\documentclass[9pt]{article}

\usepackage[utf8]{inputenc}
\usepackage{subfiles}

\usepackage{listings}
\usepackage{verbatim}
\usepackage{forest}
\usepackage[hidelinks]{hyperref}
\usepackage{multicol,caption}
\usepackage{geometry}
\usepackage{float}
\usepackage{amsmath}
\usepackage{amsfonts}
\usepackage{stmaryrd}

\usepackage{graphicx}
\graphicspath{{./figures/}{../figures/}}

\usepackage{biblatex} 
\usepackage{listings}
\usepackage{color}

\definecolor{dkgreen}{rgb}{0,0.6,0}
\definecolor{gray}{rgb}{0.5,0.5,0.5}
\definecolor{mauve}{rgb}{0.58,0,0.82}

\title{Verified invertible lexer using regular expressions and DFAs}
\author{Samuel Chassot\\EPFL, Switzerland\\
\href{mailto:samuel.chassot@epfl.ch}{samuel.chassot@epfl.ch}
\and
Viktor Kunčak\\EPFL, Switzerland\\
\href{mailto:viktor.kuncak@epfl.ch}{viktor.kuncak@epfl.ch}
}

\date{}

\begin{document}


\maketitle


\section{Introduction}
In this project, we explore the concept of invertibility applied to serialisation and lexing frameworks. 

Recall that, on one hand, serialisation is the process of taking a data structure and writing it to a bit array while parsing is the reverse operation, i.e., reading the bit array and constructing the data structure back. While lexing, on the other hand, is the process of reading a stream of characters and splitting them into tokens, by following a list of given rules. While used in different applications, both are similar in their abstract operation: they both take a list of simple characters and extract a more complex structure.

Applications in which these two operations are used are different but they share a need for the invertibility of the process. For example, when tokenising a code file that was prettyprinted by a compiler, one would expect to get the same sequence of tokens. Similarly, when a spacecraft sends scientific data to the ground, one would expect the parsed data to be the same as the one serialised by the spacecraft.

The idea of this project is to explore the idea of having a framework capable of generating parser/serialiser or lexer/prettyprinter pairs with a formally verified notion of invertibility.

We first explore related works and frameworks. After that, we present our verified lexer framework developed in Scala and verified using the Stainless framework\footnote{\url{https://github.com/epfl-lara/stainless}}. We explain the implementation choices we make and present the specifications and their proofs.

The code of the lexer with the proofs is available on Github\footnote{\url{https://github.com/samuelchassot/VerifiedLexer}}. The \textit{main} branch contains the regular expression (called regex from now on) matcher version and the verified Computable Languages while the \textit{dfa\_match} branch contains the version using the DFA matcher.

\section{Contributions}
In this project, we 
\begin{itemize}
    \item provide an overview of some of the existing works around parsing/lexing and formal verification with hands-on experiment;
    \item develop a Scala implementation of a regular expression matcher formally verified with Stainless;
    \item develop a Scala implementation of a DFA matcher formally verified with Stainless;
    \item develop a lexer in Scala, formally verified with Stainless to be correct with respect to the maximal munch principle;
    \item explore the invertibility of the lexer with respect to a printer by proposing some conditions to ensure it and propose formally verified proofs for two of them.
\end{itemize}

\section{Related work}

\subsection{Everparse}
Everparse\cite{ramananandro2019everparse} is a framework developed by Microsoft Research to generate verified parser and serialiser pairs for binary data types. It is developed as a part of the Everest project\footnote{\url{https://www.microsoft.com/en-us/research/project/project-everest-verified-secure-implementations-https-ecosystem/}}, which aims at verifying a stack to improve the security of HTTPS.

The framework is mainly composed of 3 parts: 
\begin{itemize}
    \item \textbf{Lowparse}: a library in F* and Low* of basic parsers that are then combined to generate complex parsers for given types.\\
    \item \textbf{QuackyDucky}: a frontend that generates parser/serialiser pairs from C-like type descriptions.\\
    \item \textbf{3d}: another frontend that generates validator functions for given types written in a DSL.
\end{itemize}

They propose that valid parsers must be \textit{correct}, \textit{exact}, and \textit{bijective}. A parser is \textit{correct} when it produces the data back when called on the serialised value. It is \textit{exact} when it accepts only serialised data as input, i.e., values returned by its serialiser counterpart. Finally, a parser is \textit{bjective} if it is \textit{non-malleable} (or \textit{injective}) and \textit{complete} (or \textit{sujective}) which means that each bit stream has at most one parsed representation and that it accepts at least one bit stream representation for each possible data.

To construct the parsers and serialisers, Everparse relies on \textit{Lowparse}, a library of basic parsers that are verified once and for all. Everparse then uses combinators of parsers, which are high-order functions on parsers, to combine them to create the parser for a given data type. These combinators are also verified once and for all and thus, the only part that is verified after the parser creation is the way these combinators are arranged.

The frontend generates the code and the specification that express the conditions on the parsers. This way, the verification happens after this step and the produced code is then guaranteed to be correct with respect to this specification without relying on the compiler's correctness.

The generated code is F* and/or Low* (depending on the passed options) and can be compiled to C or OCaml. Low* is a subset of F* that can be compiled into efficient C code. The user decides whether the tool should produce high or low-level code depending on the application. The high-level code output in F*, when compiled to C, gives garbage-collected code. For efficient, in-place, operations, one should use the low-level code, which is output in Low* and can be compiled into efficient C code.

\subsubsection{Experiment}
We decide to test the framework to assess its capabilities and ease of use.

The Everparse framework is developed in F* and offers 2 tools to the user: \textit{QuackyDucky} and \textit{3d}.

\textit{QuackyDucky} is a frontend that takes as input a data type description in a C-like language and produces serialiser/parser functions in F* and/or Low* with their specification and proofs.

\textit{3d} is another frontend that takes as input a type description in a specific DSL (called 3d, which stands for "Dependent Data Description") and produces validators. These validators are functions that take as input a bit stream and check whether it is a valid serialisation according to the type description or not. The type description can also be augmented by adding \textit{actions} that can add parsing instructions to the validator function which can then do some of the parsing work.

These two tools are used in different scenarios. If one needs to deal with binary formatted data from an existing C application, \textit{QuackyDucky} is more adapted. To add verified parsers to projects in F*, \textit{3d} is the better tool.

We focus our attention on \textit{QuackyDucky} during our experimentation with the framework and we make interesting observations on use cases. 

We experience difficulties due to a lack of documentation and examples while testing this framework. We however manage to generate some basic examples with the tool. We then open a pull request on their repository with some added documentation, some examples, and a modified Docker image.

We observe that the code output by \textit{QuackyDucky} is used in a different way depending on the complexity of the type we give. For simple types, once compiled to Low* and then C, the code consists of direct \textit{read} and \textit{write} functions to read and write data from and to a buffer. For complex types (like \textit{struct} contains a variable length array) however, the produced code contains \textit{validator}, \textit{accessor}, and \textit{jumper} functions that then need to be combined by the programmer to achieve the desired behaviour. These functions are used to validate the buffer, access data in the buffer at a given position, and get the position of a given field in the buffer. This means that the produced code is not a single parser function in itself but rather a set of helper functions. This is done to provide efficient code that works in place on the buffer. The user then uses these functions to access the needed values only, without necessarily parsing the entire object.

Everparse thus does not provide what we were looking for, as it does not produce pairs of parser and serialiser functions directly. We however understand the philosophy behind Everparse and why it is built that way given the role it is fulfilling in the Everest project. 

\subsection{Coqlex}
\textit{Coqlex}\cite{ouedraogo2021demo} is a framework that can generate a verified lexer based on a description of the rules. The produced lexer is guaranteed to be correct with respect to a given specification.

A lexer, according to their definition, is a function that takes as input a string to analyse, a set of rules, and a start position. Each rule is composed of a regex and a semantic action that creates a token. The lexer then returns a token, the remaining string, and the position after the analysis. 

The specification for the lexer they propose is the \textit{longest match rule}, also called \textit{maximal munch principle} (this principle is explained in detail in the section \ref{implementation:lexer:specs:maxMunch}). This principle states that the lexer must return the token produced by the rule whose regex is matching the longest prefix of the input string. In case of a tie, the first rule in the list wins (\textit{priority rule}).

Coqlex then produces a lexer implementation in Coq that is verified to follow that specification given a list of rules. These are given in a configuration file in a similar syntax to OCamllex files.

We do not explore this framework in more detail as it appears to us to have similar functionality as Verbatim (section \ref{verbatim}) which is a more mature project. For this reason, we focus on Verbatim.

\subsection{Verbatim}
\label{verbatim}
Verbatim\cite{egolf2021verbatim} is a verified lexer implemented and verified in Coq. It is based on the Brzozowski derivatives and is verified to respect the \textit{maximal munch principle} (see \ref{implementation:lexer:specs:maxMunch}). The concept of Brzozowski derivatives is detailed in the section \ref{background:regex:derivative} and the \textit{maximal munch principle} is explained in the section \ref{implementation:lexer:specs:maxMunch}.

Because the Brzozowski derivatives method has bad performance, Verbatim has an extension, called Verbatim++\cite{egolf2022verbatim++} to improve it. Verbatim++ uses a translation from regex to DFA to reach that goal. 

In a nutshell, the main idea they use to construct a DFA for a given regex is to use a 2D array to represent the transition function and the set of all possible regexes as the set of states of the DFA. The columns of the transitions array are labeled by the elements of the alphabet and the rows by the regexes. Each element is then the derivative of the regex with respect to the element of the alphabet. This array is filled by a recursive procedure. This idea was given by Brzozowski\cite{brzozowski1964derivatives}.

Verbatim++ also offers the possibility to semantically enrich tokens by letting them carry values of types defined by the user.

The lexer produced by Verbatim is verified to satisfy its maximal munch specification. It however does not explore anything related to an invertibility property.

\subsubsection{Experiment}
We try the Verbatim tool to assess its ease of use and its capabilities.

First, the version of the tool developed for the first Verbatim paper\cite{egolf2021verbatim} works with an old version of Coq and is therefore not the preferred way to use the tool. The version for the Verbatim++ paper\cite{egolf2022verbatim++} however works well with the latest version of Coq. We create a lexer for a basic LISP-like language. The code can be found on a fork of the repository\footnote{\url{https://github.com/samuelchassot/Verbatim/tree/main/Examples/BasicLisp}}. 

The configuration is given in a format close to Coq syntax. The user provides the definitions of her rules, containing a regex and a label. The regexes are constructed using the provided primitives. The user can also give semantic transformation for the tokens in the following way: she first has to give a mapping from tokens' labels to type; then she has to write a function that returns a value of this type for a given token as well as a proof that this function does not change the label of the token.

The documentation is easy to follow and the provided examples are complete enough for an easy hands-on experience.

\section{Implementation}
We now describe the implementation of the different parts of our framework and the choices we make.
\subsection{Regex matcher}
\label{implementation:regexMatcher}
We start by implementing a regex matcher. The regexes are defined for characters of a generic type. This way, our matcher can be used with any alphabet as long as the characters are of a defined type. In the following, we will call one element of the alphabet a \textit{character}, which is not necessarily of the type \textit{Character} (i.e., one byte). We also call a \textit{string} a list of such characters even though it is not an instance of the type \textit{String}.

The regexes are implemented as the following algebraic data types:

\begin{lstlisting}
abstract sealed class Regex[C]
case class ElementMatch[C](c: C) extends Regex[C]
case class Star[C](reg: Regex[C]) extends Regex[C]
case class Union[C](regOne: Regex[C], regTwo: Regex[C]) extends Regex[C]
case class Concat[C](regOne: Regex[C], regTwo: Regex[C]) extends Regex[C]
case class EmptyExpr[C]() extends Regex[C]
case class EmptyLang[C]() extends Regex[C]
\end{lstlisting}

The different primitives are the ones described in the section \ref{background:regex}. Their respective usage is the following (we indicate in parenthesis the equivalent theoretical notation used in the section \ref{background:regex}):
\begin{enumerate}
    \item \textit{ElementMatch} represents a regex matching one character ($r = \llbracket a \rrbracket$).
    \item \textit{Star} is the Kleene star operator: the regex matches any string which is composed of zero or more strings matched by the inner regex ($r = r^*$).
   \item \textit{Union} returns a regex matching any strings matched by either of the two inner regexes ($r = r_1 + r_2$).
   \item \textit{Concat} returns a regex matching any string of the format $s_1 :: s_2$ where $s_1$ is matched by the first inner regex and $s_2$ is matched by the second inner regex ($r = r_1 \cdot r_2$).
   \item \textit{EmptyExpr} is the regex which matches only the empty string ($r = \epsilon$).
   \item \textit{EmptyLang} is the regex which matches no string ($r = \emptyset$).
\end{enumerate}

The matcher uses the Brzozowski derivatives technique. The match function \textit{matchR} is implemented recursively as follows:
\begin{lstlisting}
def matchR[C](r: Regex[C], input: List[C]): Boolean = {
  if (input.isEmpty) nullable(r) else matchR(derivativeStep(r, input.head), input.tail)
} 

def derivativeStep[C](r: Regex[C], a: C): Regex[C] = {
  r match {
    case EmptyExpr()       => EmptyLang()
    case EmptyLang()       => EmptyLang()
    case ElementMatch(c)   => if (a == c) EmptyExpr() else EmptyLang()
    case Union(rOne, rTwo) => Union(derivativeStep(rOne, a), derivativeStep(rTwo, a))
    case Star(rInner)      => Concat(derivativeStep(rInner, a), Star(rInner))
    case Concat(rOne, rTwo) => {
      if (nullable(rOne)) Union(Concat(derivativeStep(rOne, a), rTwo), derivativeStep(rTwo, a))
      else Union(Concat(derivativeStep(rOne, a), rTwo), EmptyLang())
    }
  }
}
    
def derivative[C](r: Regex[C], input: List[C]): Regex[C] = {
  input match {
    case Cons(hd, tl) => derivative(derivativeStep(r, hd), tl)
    case Nil()        => r
    }
}
\end{lstlisting}

The \textit{matchR} function returns a boolean given an input string. It computes the derivatives recursively for all the input's characters and checks the nullability of the resulting regex.

To improve the performance of the lexer, we also implement a function that returns the longest matched prefix for a given input string and regex. This avoids calling \textit{matchR} on every prefix of the string and thus reduces the number of derivatives it computes. This function has the following signature:

\begin{lstlisting}
def findLongestMatch[C](r: Regex[C], input: List[C]): (List[C], List[C]) = { // ... }
\end{lstlisting}

The \textit{findLongestMatch} function consumes the input string while computing the derivatives and returns the longest matched prefix along with the suffix. In this way, the derivatives are computed once for each input character, instead of $N$ times.

\subsubsection{Specification and proofs}
\label{implementation:regex:specificationProofs}
In this section, we detail the specification of the regex matcher and a sketch of the proof.

We first look at the \textit{matchR} function and its specification. These lemmas are the matching rules described in the section \ref{background:regex} and represent the specification of the regex matcher:

\begin{lstlisting}
// EmptyString Lemma
def lemmaRegexEmptyStringAcceptsTheEmptyString[C](r: EmptyExpr[C]): Unit = {
  require(validRegex(r))
  // ...
} ensuring (matchR(r, List()))

// Single Character Lemma
def lemmaElementRegexAcceptsItsCharacterAndOnlyIt[C](r: ElementMatch[C], c: C, d: C): Unit = {
  require(validRegex(r) && r == ElementMatch(c))
  require(c != d)
  // ...
} ensuring (matchR(r, List(c)) && !matchR(r, List(d)))

// Union lemmas
def lemmaRegexAcceptsStringThenUnionWithAnotherAcceptsToo[C](r1: Regex[C], r2: Regex[C], 
  s: List[C]): Unit = {
  require(validRegex(r1) && validRegex(r2))
  require(matchR(r1, s))
  // ...
} ensuring (matchR(Union(r1, r2), s))

def lemmaRegexUnionAcceptsThenOneOfTheTwoAccepts[C](r1: Regex[C], r2: Regex[C], s: List[C]): Unit = {
  require(validRegex(r1) && validRegex(r2))
  require(matchR(Union(r1, r2), s)) 
  // ...
} ensuring (matchR(r1, s) || matchR(r2, s))

// Concat lemmas
def lemmaTwoRegexMatchThenConcatMatchesConcatString[C](r1: Regex[C], r2: Regex[C], s1: List[C], s2: List[C]): Unit = {
  require(validRegex(r1) && validRegex(r2))
  require(matchR(r1, s1))
  require(matchR(r2, s2))
  // ...
} ensuring (matchR(Concat(r1, r2), s1 ++ s2))

// Star lemmas
def lemmaStarAcceptsEmptyString[C](r: Star[C]): Unit = {
  require(validRegex(r))
  // ...
} ensuring (matchR(r, List()))

def lemmaStarApp[C](r: Regex[C], s1: List[C], s2: List[C]): Unit = {
  require(validRegex(r))
  require(matchR(r, s1))
  require(matchR(Star(r), s2))
  // ...
} ensuring (matchR(Star(r), s1 ++ s2))
\end{lstlisting}

The proofs are mainly done by induction.

Next, we define and verify the \textit{findLongestMatch} specification against the \textit{matchR} function. We indeed want that:
\begin{enumerate}
    \item the string returned by \textit{findLongestMatch} is a prefix of the input string,
    \item the string returned by \textit{findLongestMatch} is matched by the \textit{matchR} function,
    \item no longer prefix of the input string can be matched by \textit{matchR}.
\end{enumerate}

The proof of this specification is also done by induction but is more complicated than the \textit{matchR} specification and requires more intermediate lemmas.

The regex \textit{validity} is defined as a function, even though the specification it encodes is enforced by the compiler:
\begin{lstlisting}
 def validRegex[C](r: Regex[C]): Boolean = r match {
    case ElementMatch(c)    => true
    case Star(r)            => validRegex(r)
    case Union(rOne, rTwo)  => validRegex(rOne) && validRegex(rTwo)
    case Concat(rOne, rTwo) => validRegex(rOne) && validRegex(rTwo)
    case EmptyExpr()        => true
    case EmptyLang()        => true
  }
\end{lstlisting}

For the lexer invertibility proof, we also need to prove lemmas on the regex matcher related to the concepts of \textit{used characters} and \textit{first characters}. We define the set of \textit{used characters} of a regex as follows: 
\begin{lstlisting}
def usedCharacters[C](r: Regex[C]): List[C] = {
    r match {
      case EmptyExpr()        => Nil[C]()
      case EmptyLang()        => Nil[C]()
      case ElementMatch(c)    => List(c)
      case Star(r)            => usedCharacters(r)
      case Union(rOne, rTwo)  => usedCharacters(rOne) ++ usedCharacters(rTwo)
      case Concat(rOne, rTwo) => usedCharacters(rOne) ++ usedCharacters(rTwo)
    }
  }
\end{lstlisting}
The \textit{used characters} of a regex are therefore the characters that any string matched by this regex contains.

We then define the set of \textit{first characters} as follows:
\begin{lstlisting}
def firstChars[C](r: Regex[C]): List[C] = {
    r match {
      case EmptyExpr()                           => Nil[C]()
      case EmptyLang()                           => Nil[C]()
      case ElementMatch(c)                       => List(c)
      case Star(r)                               => firstChars(r)
      case Union(rOne, rTwo)                     => firstChars(rOne) ++ firstChars(rTwo)
      case Concat(rOne, rTwo) if nullable(rOne)  => firstChars(rOne) ++ firstChars(rTwo)
      case Concat(rOne, rTwo) if !nullable(rOne) => firstChars(rOne)
    }
  }
\end{lstlisting}
The \textit{first characters} of a regex are the characters that any string matched by this regex starts with.

We also know that:
\begin{align}
    firstChars(r) \subseteq usedChars(r)
\end{align}

The main lemmas about the used characters and first characters are the following:
\begin{lstlisting}
def lemmaRegexCannotMatchAStringContainingACharItDoesNotContain[C](r: Regex[C], s: List[C], 
  c: C): Unit = {
    require(validRegex(r))
    require(s.contains(c))
    require(!usedCharacters(r).contains(c))

} ensuring (!matchR(r, s))

def lemmaRegexCannotMatchAStringStartingWithACharItDoesNotContain[C](r: Regex[C], s: List[C], 
  c: C): Unit = {
    require(validRegex(r))
    require(s.contains(c))
    require(s.head == c)
    require(!usedCharacters(r).contains(c))

} ensuring (!matchR(r, s))
\end{lstlisting}

They are both proved by induction. In a nutshell, they say that a regex cannot match a string that contains (respectively starts with) a character that is not in its \textit{used characters} set (respectively in its \textit{first characters} set).

One interesting part of the proof is that, in the case of a regex of the type \textit{Concat}, the induction requires splitting the input string into the two substrings that are matched by each of the inner regexes. This requires a function that computes this separation and some lemmas to express how it relates to the \textit{matchR} function.

\subsection{Optimisations}
We implement and verify two main optimizations to the regex matching algorithm. Indeed, the runtime complexity of the Brzozowski derivatives is high (at least $O(n^3)$ based on our initial benchmark analysis). The two implementations we implement are memoization and a new representation, the zippers.

\subsubsection{Memoisation}
One way to improve the runtime performance of our derivative based matching algorithm is to use memoisation not to recompute the same derivative multiple times.

We use a Hash Table implementation with generically typed keys to implement the memoization table. We implement and formally verify this Hash Table building on top of an existing verified implementation of a Hash Table with 64-bits integer keys\cite{10.1007/978-3-031-63498-7_18}. 

This Hash Table, being a cache for the \verb|derivativeStep[C](r: Regex[C], c: C): Regex[C]|, must satisfy an invariant of the following form:
$$
\forall (r, c), cache((r, c)) = some(rd) \implies rd = derivativeStep(r, c)
$$

We use lemmas offered by our Hash Table implementation to prove the preservation of this invariant and implement a memoized version of the function: \verb|derivativeStepMem[C](r: Regex[C], c: C)(cache: Cache[C]): Regex[C]| where \verb|Cache[C]| is using the mentioned Hash Table, while offering some helper functions.

We prove that this memoized function is equivalent, i.e., it produces the same result for the same inputs as the original function. Therefore, they can be used interchangeably.

\subsubsection{Zipper}
The other optimization we implement and verify is a new representation for regex introduced by Edelmann\cite{edelmann2021efficient}. The derivation procedure is adapted for this structure.

Edelmann provides a Coq implementation and proof of the zipper-based implementation.

We implement this procedure in Scala and prove its correctness using Stainless again. The specification is again defined in terms of equivalence with the original regex matching function:
$$
\forall r, s, matchR(r, s) = matchZipper(toZipper(r), s)
$$

where \verb|toZipper| is a function that builds a corresponding zipper for a given regex.

Our preliminary performance analysis shows that the complexity goes from $O(n^3)$ down to $O(n)$, at least in analyzed cases.

\subsection{NFA detour}
Regexes are widely used and easy to use for developers but the matching algorithm using the derivatives is computationally expensive. For this reason, we explore a string-matching technique using Non-deterministic Finite Automata (NFA). As stated by Kleene's theorem, it exists an equivalent NFA for every regex. Matching a string against an automaton is computationally less expensive than against regexes.

To construct an NFA from a regex, we use the procedure proposed by R. Edelmann\cite{edelmann2021efficient}. The proposed procedure is implemented in a continuation-passing style which makes it concise and elegant.

The non-determinism nature of NFAs or more specifically the unbound number of empty transitions and possible empty transition loops make the simulation function difficult to work with to write formally verified proofs. First of all, the termination of the function is non-trivial and difficult to prove. Moreover, we need to prove the correctness of the regex to NFA by proving the two match functions are equivalent.

As writing the proof in Stainless and as the result is already known and formally proven, we decide to assume the equivalence between the input regex and the corresponding NFA. For this reason, it makes the version using NFA less interesting in our opinion and that is why we do not.

\subsection{Computable Language detour}
We explore another way of representing languages than regex with these \textit{Computable Languages}\footnote{\url{https://github.com/epfl-lara/bolts/blob/master/WIP/computable-languages/ComputableLanguage.scala}}. The implementation is an existing implementation part of the Bolts repository\footnote{\url{https://github.com/epfl-lara/bolts}} (containing scala code verified with Stainless) which is a collection of software verified with the Stainless framework. 

A language is defined by a function taking a list of characters of a generic type and returning a boolean. This function is a \textit{contains} function for the language. Then, other high-order functions are defined to construct the concatenation, the complement, the union, the star closure, and the n-power of languages.

\subsubsection{Specification and proofs}

We provide proofs for the same specification that we defined for the regexes in the section \ref{implementation:regex:specificationProofs} for this \textit{Computable Lanugages} implementation. While verifying this specification, we discover a bug in the original implementation, demonstrating the usefulness of formal verification. The patch and the proofs have been merged.

This matcher could be used in the lexer after some refactoring. We do not do it, however, we think this detour is interesting.

\subsection{DFA matcher}
\label{implementation:dfaMatcher}
Because the regex matcher has bad performance, and to solve the problems the NFA simulation poses, we implement a DFA matcher. Unlike what we do with the NFA matcher, the user directly inputs DFAs instead of the framework performing a pre-processing step to create them from regexes.

Here is the algebraic data type definition of our DFA in Scala:

\begin{lstlisting}
case class State(label: BigInt) {
  require(label >= 0)
}
case class Transition[C](from: State, c: C, to: State)
case class DFA[C](startState: State, finalStates: List[State], errorState: State, transitions: List[Transition[C]])
\end{lstlisting}

We define some invariant for DFAs to be considered as \textit{valid} DFAs: 
\begin{lstlisting}
def validDFA[C](dfa: DFA[C]): Boolean =
  uniqueStateCharTransitions(dfa.transitions, Nil()) && 
    noTransitionOutOfErrorState(dfa.transitions, dfa.errorState) && !dfa.finalStates.contains(dfa.errorState)
\end{lstlisting}

This invariant ensures that:
\begin{itemize}
    \item there exist no two transitions labeled with the same character going out of any state,
    \item the \textit{errorState} is trap state, i.e, there exists no transition going out of it,
    \item the \textit{errorState} is not a final state (i.e., accepting state).
\end{itemize}

We define the \textit{findLongestMatch} function which returns the longest prefix of a given string matched by the DFA (i.e., ending on a final state when simulated on the DFA). We then define the \textit{matchDFA} function using the \textit{findLongestMatch}: a DFA matches a given string if \textit{findLongestMatch} returns the entire string.

To improve the performance, the \textit{findLongestMatch} uses the fact that \textit{errorState} is a trap state and returns \textit{Nil} once it reaches it without continuing to explore the DFA.

\subsubsection{Specification and proofs}
\label{implementation:dfaMatcher:specs}
As the users directly input DFA, we do not prove equivalence with regexes. We also do not prove that our \textit{matchDFA} and \textit{findLongestMatch} functions follow some specification as our implementation is the specification.

We however provide proofs for the same properties that we verify for regexes about the \textit{used characters} and the fact that a DFA cannot match a string that contains a character not present in its \textit{used characters}. This has to be proven to use DFA as a direct replacement for regexes in our lexer (we explore this in the section \ref{implementation:lexer:specs:invertability}). The proof is mainly done by induction and needs lemmas that prove, for example, the relationship between DFA traversals starting at different states with different prefixes of the original tested string already consumed.

\subsection{Lexer}
\label{implementation:lexer}
In this section, we explore the implementation of the lexer.

First, here are the Scala algebraic datatypes of the \textit{rules} and \textit{tokens}:
\begin{lstlisting}
case class Rule[C](regex: Regex[C], tag: String, isSeparator: Boolean)
case class Token[C](characters: List[C], tag: String, isSeparator: Boolean)
\end{lstlisting}

The attribute \textit{isSeparator} is discussed later during the specification discussion in the section \ref{implementation:lexer:specs: invertibility:fromTokens}.

The lexer's main function has the following signature:
\begin{lstlisting}
 def lex[C](rules: List[Rule[C]], input: List[C]): (List[Token[C]], List[C])
\end{lstlisting}

As described in the section \ref{background:lexer}, the \textit{lex} function takes as input a list of rules (ordered!), an input string (i.e., a list of characters), and returns the resulting list of tokens along with the remaining suffix (i.e., not tokenised suffix).

The \textit{lex} function uses the \textit{maxPrefix} function which itself uses the \textit{maxPrefixOneRule} function. Here are their respective signatures:

\begin{lstlisting}
def maxPrefix[C](rulesArg: List[Rule[C]], input: List[C]): Option[(Token[C], List[C])]
def maxPrefixOneRule[C](rule: Rule[C], input: List[C]): Option[(Token[C], List[C])]
\end{lstlisting}

The \textit{lex} function calls \textit{maxPrefix} with the input string and the list of rules. If \textit{maxPrefix} returns a token, \textit{lex} calls itself recursively on the suffix, and the produced list of tokens is returned. If the \textit{maxPrefix} function returns \textit{None}, the \textit{lex} function returns an empty list of tokens and the input string as the non-tokenised suffix.

The \textit{maxPrefix} function takes as input a list of rules and an input string. It calls \textit{maxPrefixOneRule} for each rule on the input string and returns the token which matches the longest prefix (if any).

The \textit{maxPrefixOneRule} takes as argument a rule and an input string. It calls \textit{findLongestMatch} on the regex contained in the rule and the input string: if the regex matches a prefix, the function instantiates a new token and returns it along with the remaining suffix string; if the regex matches nothing, it simply returns \textit{None}.

This algorithm is simple and produces tokenisation following the maximal munch principle.

\subsubsection{Specification and proofs - maximal munch principle}
\label{implementation:lexer:specs:maxMunch}
For the lexer to be valid (with respect to the maximal munch principle) the rules must follow some conditions that we express as an invariant (it is part of the pre and post-conditions of the various functions). This invariant states that:
\begin{itemize}
    \item each rule is a valid rule, namely its regex is valid, its regex is not nullable and the tag is not empty,
    \item no two rules have the same tag.
\end{itemize}

Note that the regex validity is defined in the section \ref{implementation:regex:specificationProofs}.

The functions \textit{lex}, \textit{maxPrefix} and \textit{maxPrefixOneRule} have post-conditions that we prove directly to help prove further lemmas and theorems more easily:
\begin{lstlisting}
def lex[C](rules: List[Rule[C]], input: List[C]): (List[Token[C]], List[C]) = {
  require(!rules.isEmpty)
  require(rulesInvariant(rules))
  // ...
} ensuring (res =>
  if (res._1.size > 0) res._2.size < input.size && !res._1.isEmpty
  else res._2 == input
)

def maxPrefix[C](rulesArg: List[Rule[C]], input: List[C]): Option[(Token[C], List[C])] = {
  require(rulesValid(rulesArg))
  require(!rulesArg.isEmpty)
  // ...
} ensuring (res => res.isEmpty || res.isDefined && (res.get._2.size < input.size && res.get._1.characters ++ res.get._2 == input))

def maxPrefixOneRule[C](rule: Rule[C], input: List[C]): Option[(Token[C], List[C])] = {
  require(ruleValid(rule))
  // ...
} ensuring (res =>
  res.isEmpty || matchR(
    rule.regex,
    res.get._1.characters
  ) && res.get._1.characters ++ res.get._2 == input && res.get._2.size < input.size && res.get._1.tag == rule.tag && res.get._1.isSeparator == rule.isSeparator
)
\end{lstlisting}

In words, the \textit{lex} function's post-condition ensures that, if it returns some tokens, the non-tokenised suffix is smaller than the input string. The \textit{maxPrefix} post-condition ensures that, if a token is produced, concatenating the token's characters to the suffix string is equal to the input string. The \textit{maxPrefixOneRule} post-condition ensures that if a token is produced, the \textit{tag} and \textit{isSeparator} properties are the same in the token and the rule, that concatenating the token's characters and the suffix is equal to the input string and that the rule's regex indeed matches the token's characters.

The maximal munch is expressed in Stainless as the following theorem:
\begin{lstlisting}
def theoremLexSoundFirstChar[C](rules: List[Rule[C]], input: List[C], suffix: List[C], tokens: List[Token[C]], r: Rule[C], otherR: Rule[C], otherP: List[C]): Unit = {
  require(!rules.isEmpty)
  require(rulesInvariant(rules))
  require(rules.contains(r))
  require(rules.contains(otherR))
  require(lex(rules, input) == (tokens, suffix))

  require(tokens.isEmpty || tokens.head.characters.size <= otherP.size)
  require(tokens.isEmpty || tokens.head.tag == r.tag)
  require(tokens.isEmpty || tokens.head.isSeparator == r.isSeparator)
  require(ListUtils.isPrefix(otherP, input))
  require(r != otherR)
  require({
    lemmaRuleInListAndRulesValidThenRuleIsValid(r, rules)
    tokens.isEmpty || matchR(r.regex, tokens.head.characters)
  })
  // ...
} ensuring (if (ListUtils.getIndex(rules, otherR) < ListUtils.getIndex(rules, r)) !matchR(otherR.regex, otherP)
            else tokens.size > 0 && otherP.size <= tokens.head.characters.size || !matchR(otherR.regex, otherP))
\end{lstlisting}

This theorem states that, for a given tokenisation, given that either it is empty (no tokens are produced) or the first token $t = (z, l)$ is produced by a given rule $r = (e, l)$, then no rules with a smaller index match another prefix $z'$ with $z.size <= z'.size$ and no other rule matches a prefix $z''$ with $z.size < z''.size$. In other words, no rules match a longer prefix than the one produced, and no other rule with a smaller index matches the produced prefix. This theorem proves that the \textit{lex} function follows the maximal munch specification for the first produced token, and given the recursive nature of the lexer, proves that it follows this specification for all tokens. The proof is done by induction with a set of intermediary lemmas about lists and the \textit{maxPrefix} and \textit{maxPrefixOneRule} functions.

\subsubsection{Specification and proofs - invertability}
\label{implementation:lexer:specs:invertability}
We define the notion of \textit{invertability} for lexer. In general, we define invertibility with respect to a second function, which performs a "reverse" operation. In the case of the lexer, we define the \textit{print} function which transforms a list of tokens back into a list of characters. We, therefore, have the two following functions:

\begin{lstlisting}
def lex[C](rules: List[Rule[C]], input: List[C]): (List[Token[C]], List[C])
def print[C](l: List[Token[C]]): List[C]
\end{lstlisting}

To define the invertibility for these couple of functions, we first need to differentiate between two cases: 
\begin{itemize}
    \item characters $\rightarrow$ tokens $\rightarrow$ characters: $print(lex(s)) = s$
    \item tokens $\rightarrow$ characters $\rightarrow$ tokens: $lex(print(ts)) = ts$
\end{itemize}

In the first case, we start with a list of characters, which is tokenised using the \textit{lex} function and printed back to characters. In the second case, we start with a list of tokens, which is printed to characters, and this list of characters is tokenised back. In both cases, we want the output to be the same as the input.

\subsubsection{Specification and proofs - invertibility starting with characters}
\label{implementation:lexer:specs:invertibility:fromChars}
This case is the simplest of the two. The theorem which encodes it is the following:
\begin{lstlisting}
def theoremInvertFromString[C](rules: List[Rule[C]], input: List[C]): Unit = {
  require(!rules.isEmpty)
  require(rulesInvariant(rules))
  // ...
} ensuring ({
  val (tokens, suffix) = lex(rules, input)
  print(tokens) ++ suffix == input
})
\end{lstlisting}

The proof is done by induction, using the fact that the characters consumed to create each token are stored in the token. By concatenating the tokens' characters, the produced string is then the input string that is passed to \textit{lex} in the first place. The inversion procedure is then correct with no restriction on the string: in any case, if the \textit{lex} function produces some tokens, concatenating back the characters they contain and the unprocessed suffix gives the input string back.

This concludes the proof of the invertibility in this case.

\subsubsection{Specification and proofs - invertibility starting with tokens}
\label{implementation:lexer:specs:invertibility:fromTokens}
Starting with tokens is more interesting. Indeed, the list of tokens is not necessarily produced by the lexer so an interesting problem arises: consecutive tokens, when printed back, can produce a list that would not lead to the same list of tokens once tokenised. We thus need to limit the universe of the input list of tokens to make sure it would preserve the lexing invertibility. We, therefore, look for a pre-condition on a list of tokens that would make the process indeed invertible.

Our first candidate for such a condition is to define the notion of \textit{separability} for two tokens and then define that a list of tokens is invertible if each pair of consecutive tokens is separable. The \textit{separability} of tokens is defined as:
\begin{align*}
    \forall s_1 \in L(e1). \forall s_2 \in L(e_2).\ \ prefix(p, s_2) \land p \ne \varepsilon \ \rightarrow \ s_1 \cdot p \notin L(e)
\end{align*}
Two tokens are then \textit{separable} if, taking a prefix from the second token's characters and concatenating them to the first token's character produces a string not recognised by the first token's rule's regex. 

This condition is however not restrictive enough, as we find a counter-example. Let us define the three following rules:
\begin{itemize}
    \item rule1 = (regex = $\llbracket a \rrbracket + (\llbracket a \rrbracket \cdot \llbracket b \rrbracket \cdot \llbracket c \rrbracket)$ , label = "first")
    \item rule2 = (regex = $\llbracket b \rrbracket$, label = "second")
    \item rule1 = (regex = $\llbracket c \rrbracket$, label = "third")
\end{itemize}
Then, we define the following token instances:
\begin{itemize}
    \item t1 = (characters = "a", label = "first")
    \item t2 = (characters = "b", label = "second")
    \item t3 = (characters = "c", label = "third")
\end{itemize}

The list of tokens l = (t1, t2, t3) would then be printed as the string "abc". However, the string "abc" would be lexed to the following list of tokens according to the maximal munch principle: ((characters = "abc", label = "first")) $\neq$ l. We can then see that this condition is not enough to ensure that the list of tokens would be lexed back to itself after having been printed.

As another candidate, we construct another, more restrictive condition. Let us define:
\begin{itemize}
    \item a sequence of tokens: $t_1, t_2, \dots, t_n$
    \item the corresponding list of characters of the tokens: $S_1, S_2, \dots, S_n$
    \item the corresponding sequence of regexes contained in the rules used to tokenise this list: $r_1, r_2, \dots, r_m$
\end{itemize}
We want no prefix longer than $S_1$ to match any regexes. So we define \textit{separable} as follows:
\begin{align*}
    &separable(r_i, r_j) = r_i\cdot r_j \cap B = \emptyset\\
    &B = \bigcup_{k_1}^m prefixSet(r_k)
\end{align*}
where \textit{prefixSet} is a function that computes all the prefixes of all the strings contained in the language denoted by $r$.

We claim that this condition is sufficient to ensure the invertibility of the lexer. We however also claim that it is unnecessarily complicated for the applications of such an invertible lexer. The programmer who uses this lexer would indeed have to ensure that the list of tokens passed to the \textit{print} function complies with this condition to make sure that the process is invertible. This condition would be complicated to work with, mainly because of the \textit{prefixSet} function whose effects are difficult to visualise in practice. We thus claim that, even though interesting, this condition is too complicated to use in practice and therefore is not worth the time it would take to write a complete proof.

We thus propose a condition simpler to work with and sufficient to prove the invertibility based on the concept of \textit{used characters} of regexes. We separate rules (and subsequently tokens) into two groups: \textit{separators} and \textit{non-separators}. We then enforce that rules from each group use disjoint sets of characters, or formally:
\begin{align*}
    \forall r_i \in Rules_{sep}\quad \forall r_j \in Rules_{nonSep}\quad usedCharacters(r_i) \cap usedCharacters(r_j) = \emptyset
\end{align*}

The condition on the list of tokens is that tokens in the list are \textit{non-separators}, and that the special \textit{printWithSeparatorTokens} function is used to print the tokens back to characters. This function prints back the characters of the tokens but inserts between each token the given \textit{separator} token. The theorem and the print function are:
\begin{lstlisting}
def theoremInvertFromTokensSepTokenBetweenEach[C](rules: List[Rule[C]], tokens: List[Token[C]], separatorToken: Token[C]): Unit = {
  require(!rules.isEmpty)
  require(rulesInvariant(rules))
  require(rulesProduceEachTokenIndividually(rules, tokens))
  require(rulesProduceIndivualToken(rules, separatorToken))
  require(separatorToken.isSeparator)
  require(tokens.forall(!_.isSeparator))
  require({
    lemmaMaxPrefReturnTokenSoItsTagBelongsToARule(rules, separatorToken.characters, separatorToken)
    getRuleFromTag(rules, separatorToken.tag).get.isSeparator
  })
  require(sepAndNonSepRulesDisjointChars(rules, rules))
  // ...
} ensuring (lex(rules, printWithSeparatorToken(tokens, separatorToken))._1.filter(!_.isSeparator) == tokens)

def printWithSeparatorToken[C](l: List[Token[C]], separatorToken: Token[C]): List[C] = {
  require(separatorToken.isSeparator)
  l match {
    case Cons(hd, tl) => hd.characters ++ separatorToken.characters ++ 
                            printWithSeparatorToken(tl, separatorToken)
    case Nil()        => Nil[C]()
  }
}
\end{lstlisting}

The function \textit{rulesProduceEachTokenIndividually} ensures that the list of rules is capable of producing those tokens, i.e., the labels and regexes correspond to the tokens. To do so, it enforces that \textit{lex(rules, t.characters) = List(t)} for all tokens in the list.

In this version, the user gives a separator token as argument, along with a list of tokens which does not contain any separator token. The invertibility is then guaranteed modulo separator tokens which are filtered out at the end. 

The proof of this theorem relies on the proven fact that a regex cannot match a string that contains a character it does not use. As \textit{separator} and \textit{non-separator} tokens use disjoint sets of characters, if a \textit{separator} token directly follows a \textit{non-separator token}, the prefix matched by any \textit{non-separator} rule cannot be longer than the token's characters. More formally, let us have a list of \textit{non-separator} tokens 
$$t_1, \dots, t_n$$
the corresponding \textit{non-separator} rules 
$$r_1, \dots, r_n$$
and a \textit{separator} token $t_{sep}$ along with its corresponding rule $r_{sep}$. Then, no prefix of the string 
$$t_1.characters :: t_{sep}.characters :: \dots :: t_n.characters$$
longer than $t_1.characters$ can be matched by any rule in
$$r_1, \dots, r_n$$ 
because all these prefixes contain at least the character $t_{sep}.head$.

One can observe that inserting a \textit{separator} token between each pair of tokens, even though sufficient, is conservative. It is indeed not needed between every pair, as some sequence of tokens might not need a \textit{separator} token only at some specific places if any. We claim that in most applications, the \textit{separator} token insertion would be a "safety guard" more than a feature. That means that a developer would make sure that the tokens sequence is invertible without any \textit{separator} token. Then only a bug would require the print function to insert a \textit{separator} token, to ensure the preservation of invertibility.

To express this condition in our proof, we claim that the simplest way to proceed is to use the \textit{lex} function itself in the print function. The \textit{printWithSeparatorTokenWhenNeeded} function is defined as such:
\begin{lstlisting}
def printWithSeparatorTokenWhenNeeded[C](rules: List[Rule[C]], l: List[Token[C]], separatorToken: Token[C]): List[C] = {
  require(!rules.isEmpty)
  require(rulesInvariant(rules))
  require(rulesProduceEachTokenIndividually(rules, l))
  require(rulesProduceIndivualToken(rules, separatorToken))
  require(separatorToken.isSeparator)
  require(l.forall(!_.isSeparator))
  require(sepAndNonSepRulesDisjointChars(rules, rules))

  l match {
    case Cons(hd, tl) => {
      val suffix = printWithSeparatorTokenWhenNeeded(rules, tl, separatorToken)
      val maxPrefWithoutSep = maxPrefix(rules, hd.characters ++ suffix)
      maxPrefWithoutSep match {
        case Some((t, s)) if t == hd => hd.characters ++ suffix
        case Some((t, s)) if t != hd => hd.characters ++ separatorToken.characters ++ suffix
        case None() => {
          lemmaLexIsDefinedWithStrThenLexWithSuffixIsDefined(rules, hd.characters, suffix)
          check(false)
          Nil[C]()
        }
      }
    }
    case Nil() => Nil[C]()
  }
}
\end{lstlisting}

The preconditions of this function include that the rules list can produce the tokens individually, that the sets of characters used by \textit{separator} and \textit{non-separator} tokens are disjoint and that the tokens in the input list are all of the \textit{non-separator} variety.

The theorem, in this case, is the following:
\begin{lstlisting}
def theoremInvertabilityFromTokensSepTokenWhenNeeded[C](rules: List[Rule[C]], tokens: List[Token[C]], separatorToken: Token[C]): Unit = {
  require(!rules.isEmpty)
  require(rulesInvariant(rules))
  require(rulesProduceEachTokenIndividually(rules, tokens))
  require(rulesProduceIndivualToken(rules, separatorToken))
  require(separatorToken.isSeparator)
  require(tokens.forall(!_.isSeparator))
  require({
    lemmaMaxPrefReturnTokenSoItsTagBelongsToARule(rules, separatorToken.characters, separatorToken)
    getRuleFromTag(rules, separatorToken.tag).get.isSeparator
  })
  require(sepAndNonSepRulesDisjointChars(rules, rules))
// ...
} ensuring (lex(rules, printWithSeparatorTokenWhenNeeded(rules, tokens, separatorToken))._1.filter(!_.isSeparator) == tokens)
\end{lstlisting}

The post-condition of this theorem is the same as the previous one, with only the print function used which differs. The proof is however simpler since the \textit{lex} function is directly used in the \textit{printWithSeparatorTokenWhenNeeded} function to decide whether a \textit{separator} token needs to be added or not. We can indeed use the same lemmas about \textit{used characters} in the case in which a \textit{separator} token is added. In the other case, i.e., the print function does not add a \textit{separator} token, we can use the fact that the \textit{lex} function returns the same token when called on the resulting characters list because the print function tests it while printing.

\subsection{Lexer using DFA}
We implement a version of the lexer which uses the DFA matcher (see \ref{implementation:dfaMatcher}). This lexer performs the same as the version using regexes and follows the same specification. The proofs are almost identical apart from some adjustments mainly for Stainless to work with these new definitions. The DFA matcher indeed provides the same lemmas as the regex matcher.

For this version, the user needs to input the rules with DFAs directly. This makes it less intuitive to work with than the regex version but provides better performance.

\subsection{2 versions of the lexer: DFA and Regex}
The two versions of the lexer which are using regexes and DFAs respectively are completely independent. Both provide the same functionality and the same specification along with the same guarantees.

We argue that having these two independent matchers would help to add more matchers in the future. One would indeed have to prove the equivalence between her new matcher and one of ours (regex or DFA) to preserve the correctness and invertibility proofs of the lexer. Depending on the situation, one would decide which is the easiest to prove the equivalence.

It also improves usability as one of the two matchers might suit one application better than the other.

\section{Limitations and future work}
For now, our lexer only outputs simple tokens containing only the characters and a textual tag value. This is limited to be used in practice, e.g., in a compiler. Augmenting the tokens with additional values of more complex types would improve the usability of our lexer in practice.

The POSIX disambiguation lexing method\cite{posixLexing} would also be an interesting area to explore for our lexer. This could lead to performance improvement although it is only a supposition at this stage. The proof might also be interesting, for the lexer and the Stainless framework development.

Some performance improvements might be interesting to apply to our regex matcher. Applying dynamic programming techniques and caching would indeed help to improve the performance of the Brzozowski derivatives matching algorithm we currently use.

\section{Conclusion}
During this project, we explore existing work related to parsing/serialising and lexing and experiment with them in practice. We observe that the invertibility property of lexers is little studied. We, therefore, explore this property, and we observe that it cannot be true in the general case. We then propose conditions that the input list must follow to make the lexing operation invertible. We finally provide a Scala implementation of a lexer along with two different matchers: one using regular expressions and one using DFAs. We then provide proofs using the Stainless framework for the matchers and lexer correctness as well as the two invertibility conditions we propose.

\section{Acknowledgements}
I want to thank my supervisor, Viktor Kun\v{c}ak, for his valuable support and help during this project.

\newpage

\printbibliography

@InProceedings{10.1007/978-3-031-63498-7_18,
    author="Chassot, Samuel
    and Kun{\v{c}}ak, Viktor",
    editor="Benzm{\"u}ller, Christoph
    and Heule, Marijn J.H.
    and Schmidt, Renate A.",
    title="Verifying a Realistic Mutable Hash Table",
    booktitle="Automated Reasoning",
    year="2024",
    publisher="Springer Nature Switzerland",
    address="Cham",
    pages="304--314",
    isbn="978-3-031-63498-7"
}

@inproceedings{ramananandro2019everparse,
  title={$\{$EverParse$\}$: Verified Secure $\{$Zero-Copy$\}$ Parsers for Authenticated Message Formats},
  author={Ramananandro, Tahina and Delignat-Lavaud, Antoine and Fournet, C{\'e}dric and Swamy, Nikhil and Chajed, Tej and Kobeissi, Nadim and Protzenko, Jonathan},
  booktitle={28th USENIX Security Symposium (USENIX Security 19)},
  pages={1465--1482},
  year={2019}
}

@inproceedings{ouedraogo2021demo,
  title={Demo Paper: Coqlex, an approach to generate verified lexers},
  author={Ouedraogo, Wendlasida and Ilik, Danko and Stra{\ss}burger, Lutz},
  booktitle={ML 2021-ACM SIGPLAN Workshop on ML},
  year={2021}
}

@inproceedings{egolf2021verbatim,
  title={Verbatim: A Verified Lexer Generator},
  author={Egolf, Derek and Lasser, Sam and Fisher, Kathleen},
  booktitle={2021 IEEE Security and Privacy Workshops (SPW)},
  pages={92--100},
  year={2021},
  organization={IEEE}
}

@inproceedings{egolf2022verbatim++,
  title={Verbatim++: verified, optimized, and semantically rich lexing with derivatives},
  author={Egolf, Derek and Lasser, Sam and Fisher, Kathleen},
  booktitle={Proceedings of the 11th ACM SIGPLAN International Conference on Certified Programs and Proofs},
  pages={27--39},
  year={2022}
}

@article{brzozowski1964derivatives,
  title={Derivatives of regular expressions},
  author={Brzozowski, Janusz A},
  journal={Journal of the ACM (JACM)},
  volume={11},
  number={4},
  pages={481--494},
  year={1964},
  publisher={ACM New York, NY, USA}
}

@techreport{edelmann2021efficient,
  title={Efficient Parsing with Derivatives and Zippers},
  author={Edelmann, Romain},
  year={2021},
  institution={EPFL}
}

@InProceedings{posixLexing,
author="Ausaf, Fahad
and Dyckhoff, Roy
and Urban, Christian",
editor="Blanchette, Jasmin Christian
and Merz, Stephan",
title="POSIX Lexing with Derivatives of Regular Expressions (Proof Pearl)",
booktitle="Interactive Theorem Proving",
year="2016",
publisher="Springer International Publishing",
address="Cham",
pages="69--86",
isbn="978-3-319-43144-4"
}

\newpage
\section{Appendix}
\subsection{Regex}
\label{background:regex}
In this section, we define regular expressions (abbreviated regexes) and their properties. We use the definition of regexes as stated in the Verbatim++ paper\cite{egolf2022verbatim++}.

Regexes are used to inductively denote regular languages. Each regex denotes one language. We say that a regex $e$ matches a string $s$ if $s$ is part of the language denoted by $e$. 

We call a \textit{character} a \textit{symbol} which is part of the alphabet used by the set of regular expressions defined for the lexing process. 

We call a \textit{string} a list of such \textit{characters}, including the \textit{empty string} (denoted by $\lambda$) which is the empty list.

Let us define the notation 
\begin{align*}
    z \simeq e
\end{align*}
to express that the regex $e$ matches the string $z$.

A regex is recursively defined by the following grammar:
\begin{align*}
    e :== \emptyset \mid \epsilon  \mid  \llbracket a \rrbracket \mid e+e  \mid e\cdot e \mid  e^*
\end{align*}

and the matching rules are the following:

\begin{align*}
    &\text{MatchEmpty: } \lambda \simeq \epsilon & &\text{MatchChar: } a \simeq \llbracket a \rrbracket & &\text{MatchConcat: } \frac{z_1 \simeq e_1 \quad z_2 \simeq e_2}{z_1 :: z_2 \simeq e_1 \cdot e_2} \\
    &\text{MatchUnionL: } \frac{z \simeq e_1 }{z \simeq e_1 + e_2} & &\text{MatchUnionR: } \frac{z \simeq e_2 }{z \simeq e_1 + e_2} & \\
    &\text{MatchStarEmpty: } \lambda \simeq e^* & &\text{MatchStar: } \frac{z_1 \simeq e \quad z_2 \simeq e^*}{z_1 :: z_2 \simeq e^*} 
\end{align*}

\subsubsection{Derivative}
\label{background:regex:derivative}
The \textit{derivative} of a language is a concept first introduced by Brzozowksi \cite{brzozowski1964derivatives}. In this work, we use a string-matching algorithm for regexes that uses the Brzozowski derivatives.

The derivative is defined for a given language with respect to a given character:
\begin{align*}
    \partial_a L = \{z \mid a::z\in L\}
\end{align*}

Informally, the derivative of a language $L$ with respect to a character $a$ is the new language that is obtained by keeping only suffixes of strings starting with the character $a$.

The definition of derivative can be extended to a string recursively as follows:

\begin{align*}
    \partial_\lambda L &= L\\
    \partial_{a::z} L &= \partial_a(\partial_z L) 
\end{align*}

Then, one can observe that:
\begin{align*}
    z \in L \Leftrightarrow \lambda \in \partial_z L 
\end{align*}

This property can be used to test whether a string belongs to a language or not. To do so, one would compute the derivative of the language with respect to the given string and check whether the empty string belongs to this new language.

The derivative concept can be extended to regexes as they are denoting languages. We define $\partial_a r = r'$ where $r$ is a regex denoting the language $L$ and $r'$ denotes the language $\partial_a L$. The other definitions follow.

Here is the algorithm to recursively compute the derivative of a regex with respect to a character:
\begin{align*}
    \partial_a \emptyset &= \emptyset\\
    \partial_a \epsilon &= \emptyset\\
    \partial_a \llbracket a \rrbracket &= \epsilon\\
    \partial_a \llbracket b \rrbracket &= \emptyset\\
    \partial_a (r_1 + r_2) &= \partial_a r_1 + \partial_a r_2 \\
    \partial_a (r_1 \cdot r_2) &= (\partial_a r_1 \cdot r_2) + (\text{if nullable } r_1 \text{ then } \partial_a r_2) \text{ else } \emptyset\\
\end{align*}

\textit{Nullable} is a function defined on a regex that returns true if the empty string is part of the language denoted by the regex. It is computed as follows:

\begin{align*}
    \text{nullable}(\emptyset) &= false \\
    \text{nullable}(\epsilon) &= true  \\
    \text{nullable}(\llbracket a \rrbracket) &= false  \\
    \text{nullable}(r^*) &= true \\
    \text{nullable}(r_1 + r_2) &= \text{nullable}(r_1) \lor \text{nullable}(r_2)   \\
    \text{nullable}(r_1 \cdot r_2) &= \text{nullable}(r_1) \land \text{nullable}(r_2)  \\
\end{align*}

These can be combined to obtain a procedure checking whether a string $z$ is recognised by a regex $r$.

\subsection{DFA}
\label{background:dfa}
Deterministic Finite Automata are another way of denoting regular languages. We use the definition provided in Verbatim++\cite{egolf2022verbatim++}. 

Let us recall this definition and the basic properties we use in our project.

A DFA $M$ is a 5-tuple $$(Q, \Sigma, \delta, q_0, F)$$ with
\begin{itemize}
    \item $Q$: a finite set of states
    \item $\Sigma$: a finite alphabet
    \item $\delta$: the transition function of the form $Q \times \Sigma \rightarrow Q$
    \item $q_0 \in Q$: the start state
    \item $F \subseteq Q$: the set of accepting state
\end{itemize}

A DFA can be represented as a directed graph in which each vertex is a state in $Q$ and the edges represent the transition function. An edge labeled with $a \in \Sigma$ exists from the vertex $s_1 \in Q$ to the vertex $s_2 \in Q$ if $\delta(s_1, a) = s_2$.

To check whether a string $z = a_1a_2 \dots a_n$ with $a_i \in \Sigma$ belongs to the language defined by a DFA, we apply the following procedure: we start at the state $q_0$ and follow transitions by consuming characters of the string $z$; if no transition labeled with the next character exists, the string does not belong to the language. When all the characters in $z$ are consumed, let $s$ be the current state; if $s\in F$ then the string $z$ belongs to the language, otherwise it does not.

\subsection{NFA}
\label{background:nfa}
Let us define Non-deterministic Finite Automata, which are another type of automata that can be used to denote regular languages. NFA is a generalisation of DFA, in particular, every DFA is an NFA.

The definition of NFAs can vary, therefore we present the definition we use for this project. 

An NFA is represented by the same 5-tuple of DFAs defined in the section \ref{background:dfa} with one key difference:
\begin{itemize}
    \item $\delta$: the transition function of the form $Q \times (\Sigma \cup \epsilon) \rightarrow \mathcal{P}(Q)$
\end{itemize}
where $\mathcal{P}(Q)$ is the power set of $Q$ and $\epsilon$ the empty string.

This means that, from a given state, multiple transitions can be labeled with the same character, and a transition can be labeled with $\epsilon$, i.e., it does not consume any character when followed.

This difference makes the NFAs Non-deterministic, in the sense that, when simulated, there is not a single path to follow, multiple states can be visited at the same time.

Using the powerset construction method, one can construct a DFA which recognises the same language as any given NFA. NFAs and DFAs are therefore completely equivalent.

\subsection{Lexer}
\label{background:lexer}
In this section, we define the notions of \textit{tokens}, \textit{rules}, and \textit{lexer} we use in our project.

First, a \textit{token} is a tuple containing a sequence of characters (i.e., a string) and a label (or \textit{tag}):
\begin{align*}
    t = (z, l)
\end{align*}

Then, a \textit{rule} is a tuple containing a regular language representation (it can be a regex, a DFA, ...) and a label (or \textit{tag}):
\begin{align*}
    r = (L, l)
\end{align*}

Finally, a lexer is defined as a function that takes a list of rules and a sequence of characters as input and produces a sequence of tokens. To do so, the lexer repeatedly produces a new token by taking a prefix out of the input sequence of characters. This prefix belongs to the language of one of the rules, and the label of that rule is given to the newly created token. This process is called \textit{lexing}, or \textit{tokenisation}. The lexer also returns the remaining suffix of characters that cannot be lexed.

Such a lexer can produce different tokenisation for the same input string and list of rules, that is why we define the \textit{maximal munch principle} which ensures that the produced list of tokens is unique. This principle states that, when the lexer produces a new token, the prefix taken out of the input sequence of characters must be the longest possible that matches one of the rules. If multiple rules match the same prefix, the rule with the smallest index (i.e., the highest priority) is used to create the token. This principle ensures that the tokenisation process produces a unique output.

In this approach, the lexer first finds the longest matching prefix for every rule in the list and then constructs the new token depending on which prefix is the longest and on the priority of the rules.

Fahad Ausaf, Roy Dyckhoff, and Christian Urban explore another approach to lexing in their paper\cite{posixLexing}. The idea is to use the POSIX disambiguation to select the rule to use to create a token. To do so, the lexer first constructs a new regex $e = e_1 + e_2 + \dots + e_n$ for $e_1, \dots, e_n$ the regexes of the rules $r_1, \dots, r_n$. Then, the lexer finds the longest prefix of the input string matched by the regex $e$. The POSIX disambiguation is then used to determine which part of the regex $e$ indeed matches the string and thus determine which of the regexes $e_1, \dots, e_n$ matches the string. With this information, the lexer can then use the corresponding rule $r_i$ to create the new token. This approach is interesting and would be interesting to explore through the formal verification prism in the future.

\newpage

\begin{lstlisting}
def matchRSpec[C](r: Regex[C], s: List[C]): Boolean = {
    require(validRegex(r))
    decreases(s.size + regexDepth(r))
    r match {
      case EmptyExpr()     => s.isEmpty
      case EmptyLang()     => false
      case ElementMatch(c) => s == List(c)
      case Union(r1, r2)   => matchRSpec(r1, s) || matchRSpec(r2, s)
      case Star(rInner)    => s.isEmpty || findConcatSeparation(rInner, Star(rInner), Nil(), s, s).isDefined
      case Concat(r1, r2)  => findConcatSeparation(r1, r2, Nil(), s, s).isDefined
    }
  }
\end{lstlisting}

\begin{lstlisting}
def mainMatchTheorem[C](r: Regex[C], s: List[C]): Unit = {
    require(validRegex(r))
    decreases(s.size + regexDepth(r))
    // ...
} ensuring (_ => matchR(r, s) == matchRSpec(r, s))
\end{lstlisting}
\end{document}